\documentclass[%
12pt,T
oneocolumn,
nofootinbib,
amsmath,amssymb,
aps,
floatfix,
unsortedaddress
]{revtex4-2}

\usepackage{graphicx,color}
\usepackage{subfig}
\usepackage{dcolumn}
\usepackage{bm}
\usepackage{graphicx,color}
\usepackage{float}
\usepackage{romannum}
\usepackage{amsmath}
\usepackage{mathtools}
\usepackage{upgreek}
\usepackage{cancel}
\usepackage{multirow}
\usepackage[pdfpagelabels, pdfencoding=auto, psdextra]{hyperref}
\hypersetup{%
	pdfsubject=Paper,
	pdfkeywords={nuclear physics} {Pionless EFT} {Two-Nucleon} {Lattice} {L\"usher formula},
	unicode = true,
	breaklinks = true,
	colorlinks = true,
	linkcolor = blue,
	citecolor = blue,
	menucolor = blue,
	citecolor = blue,
	urlcolor = blue
}

\graphicspath{{./}}
\begin{document}
	\pagenumbering{arabic}
	
	\title{ Exploring $ \Lambda{\text -} $ and $ \Xi{\text -}$triton correlation functions in heavy-ion collisions }
	\author{Faisal Etminan}
	\email{fetminan@birjand.ac.ir}
	\affiliation{
		Department of Physics, Faculty of Sciences, University of Birjand, Birjand 97175-615, Iran
	}%
	\affiliation{ Interdisciplinary Theoretical and Mathematical Sciences Program (iTHEMS), RIKEN, Wako 351-0198, Japan}	
	\date{\today}%
	
	\begin{abstract}
		The $ \Lambda{\text -} $ and $ \Xi{\text -}$triton(t) momentum correlation functions, to be measured in high-energy heavy-ion collisions, are explored. 
		Mainly, STAR detector acquired data  provides an opportunity to explore the $ \Lambda t $ correlation function.
		The $ \Lambda t $  correlation functions are calculated using
		an isle-type and spin-averaged $ \Lambda t $ potential, 
		also, its sensitivity to changes in potential strength has also been investigated. 
		Besides, even though there is no experimental data on the $ \Xi{\text -} $triton interaction yet, 
		I constructed  $\Xi t$ potentials based on the first principles HAL QCD and  Nijmegen extended soft-core (ESC08c) model of spin- and isospin averaged $\Xi N$ interactions in single-folding potentials (SFP) approach.		
		Then, the  $\Xi t$ correlation functions are calculated for these two modern potentials as well as for Nijmegen hard-core model D (NHC-D) $\Xi N$ potential. 
		The numerical results predict that, with good measurement resolution, it might be possible to recognize different potentials with a correlation function at relatively small source sizes, i.e., $ R = 1-3 $ fm. 
	\end{abstract}
	
	
	\maketitle
	\section{Introduction} \label{sec:intro}
	New measurements by the STAR (Solenoidal Tracker at RHIC) Collaboration at RHIC (Relativistic Heavy Ion Collider) shows a significant increase in the $\Lambda$ binding energy of the hypertriton~\cite{AbdallahPLB2022}, in fact
	some of the past values have been called into question~\cite{BOTTA2017165}. 
	Therefore, the accurate measurements of $ \Lambda $ binding energies
	of heavier hypernuclei than the hypertriton are supposed to make progress
	our knowledge about the $\Lambda$ interactions by heavier nuclei.
	
	The $_{\Lambda}^{4}H$ nucleus, a bound state of a $\Lambda$ hyperon
	and a tritium core, is an isotope of the hydrogen. 
	It was observed in the primary helium bubble chamber and nuclear emulsion experiments~\cite{JURIC19731}.
	So far the binding energy of $_{\Lambda}^{4}H$ is measured by different
	experimental techniques such as the emulsion technique~\cite{JURIC19731},
	decay-pion spectroscopy in electron scattering~\cite{EsserPRL2015}, high-resolution
	decay-pion spectroscopy by A1 Collaboration at the Mainz
	Microtron MAMI, Germany~\cite{SchulzNPA2016} and very recently, it is measured in Au+Au
	collisions at $\sqrt{s_{NN}}=3$ GeV by the STAR Collaboration~\cite{AbdallahPLB2022,kozhevnikova2024lighthypernuclei}. 
	
	Since studying $ YN $ interactions by the classical scattering experiments is challenging, 
	recent progress in theoretical and experimental techniques in heavy-ion reactions at relativistic energies provide a unique opportunity to explore $ YN $ interactions through measuring two-particle correlations~\cite{cho2017exotic, GarridoPRC2024} and production of light hypernuclei~\cite{kozhevnikova2024lighthypernuclei}. 
	The first measurement of $ \Lambda p $ and $ \Lambda d $  correlation,
	which is done by STAR Collaborations with $\sqrt{S_{NN}}=3$ GeV Au+Au collisions~\cite{hu2023},
	shed light on both $ YN $ two-body and $ YNN $ three-body interactions~\cite{Morita2016,PhysRevC.108.064002,PhysRevX.14.031051,GarridoPRC2024},
	which is crucial for understanding neutron star properties.
	The correlation functions for some hadron-deuteron systems are explored theoretically, such as
	$ pd $~\cite{bazak2020production, PhysRevC.108.064002, mrowczynski2019hadron}, $ {K}^{-}d $~\cite{PhysRevX.14.031051, mrowczynski2019hadron},
	$ \Lambda d $~\cite{Haidenbauerprc, GarridoPRC2024}, $ \Xi d $~\cite{Ogata2021},
	and $ \Omega NN $~\cite{zhang2021production, ETMINAN2023122639}.
	Moreover very recently, the momentum correlation between $ \Lambda \alpha $~\cite{jinno2024femtoscopic},
	$ \Xi \alpha $~\cite{kamiya2024}, $ \Omega \alpha $~\cite{etminan2024omegaAlpha} and 
	$ \phi \alpha $~\cite{etminan2024phiAlpha} have been investigated theoretically.

	According to the above discussion, it is desirable to predict the $  \Lambda t$ momentum correlation functions
	numerically using phenomenological potentials to be compared by expected measuring $ \Lambda t $ correlations function by STAR in the relativistic heavy ion collisions~\cite{AbdallahPLB2022} as an independent source of information.

	In addition, I explored the formation possibility of $ \Xi{\text -}t $ system
	by evaluating its momentum correlation functions applying the effective folding $ \Xi{\text -} t $ potentials.
	However, no experimental results or observations have been reported about $ \Xi{\text -}t $ system so far,
	but evidence from femtoscopic measurements for an attractive $ \Xi^{-}p $ interaction~\cite{PhysRevLett.123.112002} have triggered significant motivation to investigate $ \Xi $ hypernuclei theoretically~\cite{LE2020135189,Hiyama2020,hiyamaPRC22}.
	Therefore. in this work, I constructed
	a three Gaussian (3G) and a  Woods-Saxon (WS) type form for $\Xi t$ potential based on the first principles HAL QCD~\cite{sasaki2020} and  Nijmegen extended soft-core (ESC08c)~\cite{nagels2015extendedsoftcore} model of $\Xi N$ interactions, respectively.
	The $\Xi t$ potentials are obtained in the single-folding potentials (SFP) approach by using the corresponding spin- and isospin averaged $\Xi N$ interactions. 
	Then the relevant $\Xi t$ momentum correlation functions are calculated by employing these two modern $\Xi t$ potentials and another potential built based on phenomenological Nijmegen hard-core model D (NHC-D) $\Xi N$ potential which is taken from the literature~\cite{myint1994}. 
	
		Since the present work is an exploratory study, it is done on a simple technical level.
	Actually, it provide an illustration for what can be expected from measuring $ Y t $ correlations.
	Indeed, in order to analyze  actual data of $ Y t $ correlations, full-fledged  and high precision
	approach of the $ Y t $ system is desired.

	The paper is structured as follows:
	In Sec.~\ref{sec:featur-of-potential}, the employed potentials in this work, i.e., the effective two-body $\Lambda t$ and $\Xi t$ potentials,  are described, and some properties are discussed. 
	The correlation function results by Koonin-Pratt (KP) formula~\cite{OHNISHI2016294} are given in Sec.~\ref{sec:Two-particle-CF} and the relevant discussions about the results also presented there.
	I summarize and  conclude my work in Sec.~\ref{sec:Summary-and-conclusions}.  
	\section{Effective two-body $\Lambda t$ and $\Xi t$ potentials   } \label{sec:featur-of-potential}	
\begin{figure*}[hbt!]
		\centering
		\includegraphics[scale=1.0]{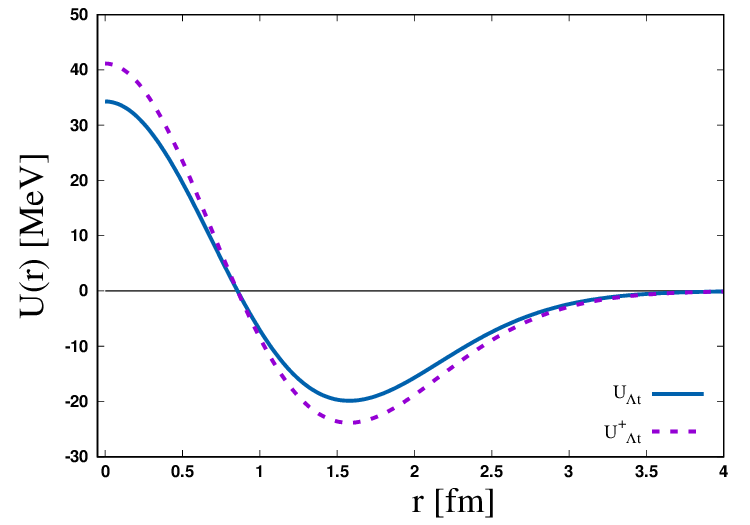}
		\caption{ The spin averaged $ \Lambda t $ potentials, $U_{\Lambda t}\left(r\right)$, as functions of distance between $ \Lambda $ and triton. 
			$U_{\Lambda t}$ (solid blue line) is given by Eq.~\eqref{eq:U_lambdat} and 
			$ U_{\Lambda t}^{+}=1.2\:U_{\Lambda t} $ (dotted violet line) simulates a $ 20 $ percent increment in $ \Lambda t $ interaction. 
			The $U_{\Lambda t}$ is taken from Ref.~\cite{myint1994}.
			\label{fig:LambT_pot}}
	\end{figure*}

	For $\Lambda t$ potential, Kurihara's isle-type potential is employed~\cite{KuriharaPRC1985}. 
	That is extracted from Dalitz's hard core $\Lambda N$
	interaction and has a two-range Gaussian form~\cite{DALITZ1972109}. 
	The strength of $ \Lambda t $ potential is tuned in such a way as to reproduce
	the experimental ground state energy of $_{\Lambda}^{4}H$, $ E=-2.04(4) $
	MeV for the singlet (S) $\Lambda t$ state~\cite{JURIC19731} and the first excited state
	energy, $ E=-0.99(4) $ MeV for the triplet (T) $ \Lambda t $ state~\cite{YamamotoPRL2015,LE2020135189}.
	 The $U_{\Lambda t}\left(r\right)$ potential between $ t $ and $ \Lambda $ is given by~\cite{myint1994}
	\begin{equation}
		U_{\Lambda t}\left(r\right)  =  \frac{1}{4}V_{\Lambda t}^{S}\left(r\right)+\frac{3}{4}V_{\Lambda t}^{T}\left(r\right)=
		\label{eq:U_lambdat}
	\end{equation}
	$$ 359.2\exp\left[-\left(\frac{r}{1.25}\right)^{2}\right]-324.9\exp\left[-\left(\frac{r}{1.41}\right)^{2}\right], $$
	this Isle potential has a two-range Gaussian form with the central repulsion and the long range attraction and it is shown in Fig.~\ref{fig:LambT_pot}.

	Furthermore, the new $ \Lambda $ binding energies measurements, performed for the $ _{\Lambda}^{4}H $ $ \left(0^{+}\right) $  
	state with respect to the $ ^{3}H+\Lambda $ mass are,
	$ 2.12\left(1\right)_{\textrm{stat.}}\left(9\right)_{\textrm{syst.}} $ MeV~\cite{EsserPRL2015}, 
	$ 2.157\left(5\right)_{\textrm{stat.}}\left(77\right)_{\textrm{syst.}} $  MeV by A1 Collaboration~\cite{SchulzNPA2016} and
	$ 2.22\left(6\right)_{\textrm{stat.}}\left(14\right)_{\textrm{syst.}} $  MeV by the STAR Collaboration~\cite{AbdallahPLB2022}.
	Since the new measurements (especially by the STAR Collaboration) show a significant
	enhancement in the $\Lambda$ binding energy of the hypertriton and $_{\Lambda}^{4}H$ hypernuclei,  
	and also to check the sensitivity of the correlation function to the nature of $ \Lambda t $ interaction,  
	the strengthened potential, $ U_{\Lambda t}^{+} $ is defined as
	\begin{equation}
		U_{\Lambda t}^{+}\left(r\right)=1.2\:U_{\Lambda t}\left(r\right). \label{eq:Uplus_lambdat}
	\end{equation}
	The strength of the original (old) potential in Eq.~\eqref{eq:U_lambdat} is increased by $ 20 $ percent. 
	To compare with $ U_{\Lambda t} $ the behavior of this potential is depicted in Fig.~\ref{fig:LambT_pot} as function of the distance between $t$ and $\Lambda$. 
	The scattering length, effective range and binding energy of the $ \Lambda t $ system
    for different potentials are summarized in Table~\ref{tab:ltriton}.
	
	\begin{table}[hbt!]
		\caption{
			Scattering length $ a_{0} $, effective range $r_{0}$ and binding energy $B_{\Lambda}$ for $ \Lambda t $ system.
			The  potential $U_{\Lambda t}\left(r\right)$ is taken directly from Ref.~\cite{myint1994}.
			\label{tab:ltriton}}	
		\begin{tabular}{ccccc}
			\hline
			\hline 
			Model &  $a_{0}$ (fm)&$r_{0}$ (fm)& $B_{\Lambda}$ (MeV)\\
			\hline
			$U_{\Lambda t}\left(r\right)$      & $5.9 $ & $2.3$ & $1.27 $ \\
			$ U_{\Lambda t}^{+}\left(r\right) $ & $4.6 $ & $2.1$ & $2.57 $ \\
			\hline
			\hline 	
		\end{tabular}
	\end{table}
	
	In the case of $\Xi t$ system, the following folding potential~\cite{Satchler1979,Etminan:2019gds}, taken from Ref.~\cite{myint1994}, has been used,
	\begin{equation}
		U_{\Xi t}\left(r\right)=\sum_{i=1}^{3}c_{i}e^{-\left(r/d_{i}\right)^{2}}, \label{eq:U_Xi-t}
	\end{equation}
with the parameters specified in Table~\ref{tab:para-XiT}. 
	$U_{\Xi t}\left(r\right)$ is built by folding an effective
	$\Xi N$ potential with the triton density distribution function $ \rho\left(r^{\prime}\right) $~\cite{myint1994,Etminan:2019gds},
	\begin{equation}
		U_{\Xi t}\left(r\right)=\int d\boldsymbol{r}^{\prime}\rho\left(r^{\prime}\right)\bar{V}_{\Xi^{-}N}\left(\boldsymbol{r}-\boldsymbol{r}^{\prime}\right),
		\label{eq:SF-method}
	\end{equation}
	where $r$ is the distance of $\Xi^{-}$ from the center-of-mass of the triton. 
	The nucleon density in the triton is evaluated from a harmonic oscillator model,
	\begin{equation}
		\rho\left(r\right)=3\left(\frac{3\beta}{2\pi}\right)^{3/2}\exp\left(-\frac{3}{2}\beta r^{2}\right), \label{eq:t-density}
	\end{equation}
	$ \beta=0.386\; \textrm{fm}^{-2} $ is the strength of harmonic oscillator
	that reproduces the matter root-mean-square (rms) radius $1.61$ fm
	which is measured in electron scattering experiments from light nuclei~\cite{CIOFIDEGLIATTI1980163}. 
	The effective potential $ \bar{V}_{\Xi^{-}N} $ is approximated from the isospin-spin averaged $\Xi^{-}N$ interaction,
	\begin{equation}
		\bar{V}_{\Xi^{-}N}=\frac{1}{8}V_{\Xi^{-}p}^{^{1}S_{0}}+\frac{3}{8}V_{\Xi^{-}p}^{^{3}S_{1}}+\frac{1}{8}V_{\Xi^{-}n}^{^{1}S_{0}}+\frac{3}{8}V_{\Xi^{-}n}^{^{3}S_{1}}, \label{eq:Xi-N-average-pot}
	\end{equation}
	the two-body potential $V_{\Xi^{-}N}$ between $\Xi^{-}$ and
	$ N $ are Shinmura's potential~\cite{myint1994} which is based on the 
	NHC-D potential\cite{PhysRevD.15.2547, PhysRevC.40.2226}. 
	The low-energy data derived with $U_{\Xi t}\left(r\right)$ indicates no bound or resonant state.
	But, if we switch on the Coulomb interaction, the bound state $ 0.14 $ MeV appears.
	 Therefore, this is a Coulomb-assisted bound state.

	In the following, the ESC08c and the HAL QCD $\Xi N$ potentials which are used to find the effective single-folding potential of $ \Xi t $ is described. 
	The  ESC08c Nijmegen $\Xi N$ potential function consists
	of local central Yukawa-type potentials with attractive and repulsive terms~\cite{nagels2015extendedsoftcore,Garcilazo2016prc}, 
	\begin{equation}
		V_{\Xi N}^{ESC08c}\left(r\right)=-A\frac{\exp\left(-\mu_{A}r\right)}{r}+B\frac{\exp\left(-\mu_{B}r\right)}{r}.
		\label{eq:pot-XiN}
	\end{equation}
	where the low-energy data and the parameters of these models are given in Table I of Ref.~\cite{Garcilazo2016prc}. 
	
	For  $ \Xi N $ HAL QCD potential, the concrete parametrizations, are taken
	straight from Ref.~\cite{sasaki2020}
	at the imaginary-time slices $ t/a=11,12,13 $ where $ a=0.0846 $ fm is the lattice spacing. Sasaki et al. in Ref.~\cite{sasaki2020} 
	clearly treated $ \Lambda\Lambda-\Xi N $ interactions  by the HAL QCD coupled-channel formalism~\cite{aoki2013,etminan2024prd}.
	The S-wave $ \Xi N $ interactions is classified in four channels $^{11}S_{0},^{31}S_{0},^{13}S_{1}$
	and $^{33}S_{1}$. Here, same as Ref.~\cite{sasaki2020}, the spectroscopic notation $^{2I+1,2S+1}S_{J}$ is employed
	where $ I, S $ and $ J $ indicate the total isospin, the total spin,
	and the total angular momentum, respectively.

	For phenomenological applications, the lattice QCD potential was fitted
	by analytic functional forms composed of multiple Gaussian and Yukawa
	functions. The Gauss functions set out the short-range part of the
	potential and  Yukawa functions describe the meson exchange picture
	at medium to long-range distances
	\begin{eqnarray}
		V_{\Xi N}\left(r\right) & =\sum_{i=1}^{3} & \alpha_{i}e^{-r^{2}/\beta_{i}^{2}}+\lambda_{2}\left[\mathcal{Y}\left(\rho_{2},m_{\pi},r\right)\right]^{2}+\lambda_{1}\mathcal{Y}\left(\rho_{1},m_{\pi},r\right),
	\end{eqnarray}
	where the values of the parameters $ \alpha_{1,2,3},\beta_{1,2,3},\lambda_{1,2} $ and
	$ \rho_{1,2} $ are given in Table 4 of Ref~\cite{sasaki2020}.
	$ m_{\pi}\simeq146 $ MeV is the pion mass that was measured on the
	lattice. 
	 The form factor $\mathcal{Y}$ defines Yukawa function 
	\begin{equation}
		\mathcal{Y}\left(\rho,m,r\right)\equiv\left(1-e^{-\frac{r^{2}}{\rho^{2}}}\right)\frac{e^{-mr}}{r}.
	\end{equation}
	Since the $ \Xi+t$ system required
	to show a clear $\Xi$-core structure, I employ approximately reliable
	expressions for spin- and isospin dependence of $\Xi N$ interaction
	to $s$-shell (the core nucleons and the $\Xi$ are in S-wave states).
	Therefore, in these approximations, the effective $\Xi N$ interactions
	can be obtained by averaging on the spin- and isospin components~\cite{hiyama2008prc,le2021} 
	\begin{equation}
		\bar{V}_{\Xi N}\simeq\frac{\left(V_{\Xi N}^{^{11}S_{0}}+3V_{\Xi N}^{^{31}S_{0}}+3V_{\Xi N}^{^{13}S_{1}}+9V_{\Xi N}^{^{33}S_{1}}\right)}{16}.\label{eq:vbar-hal}
	\end{equation}
	
\begin{figure*}[hbt!]
		\centering
			\includegraphics[scale=0.64]{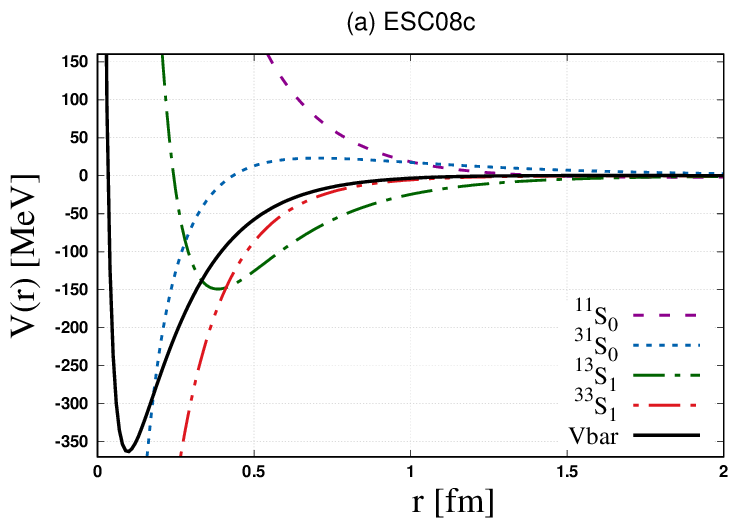} \includegraphics[scale=0.64]{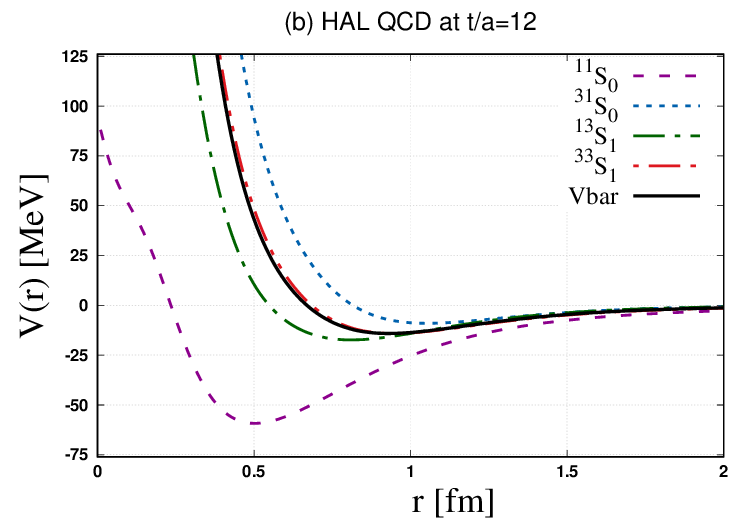}
		\caption{	(a) ESC08c  and (b) HAL QCD $ \left(t/a=12\right) $ $\Xi N $ potentials in $^{11}S_{0},^{31}S_{0},^{13}S_{1}$
			and $^{33}S_{1}$ channels. The corresponding scattering phase shifts for each channel are given in Ref.~\cite{Hiyama2020}. The solid black line, which is labeled by Vbar, shows the  spin- and isospin averaged potential, $\bar{V}_{\Xi N} $ in Eq.~\eqref{eq:vbar-hal}.	\label{fig:NXi_pot}}
	\end{figure*}

	In Fig.~\ref{fig:NXi_pot}, I compare $\Xi N$ potential for (a) the ESC08c model and (b) the HAL QCD at
	$ t/a = 12 $~\cite{sasaki2020}. The statistical errors are not shown in Fig.~\ref{fig:NXi_pot}(b), but are considered in my calculations. 

	Figure~\ref{fig:NXi_pot} reveals a qualitative difference between (a) and (b).
	Therefore, it is interesting to find out how these differences are embodied in the low energy properties of $ \Xi t $ systems.	
	
	\begin{figure*}[hbt!]
		\centering
		\includegraphics[scale=1.0]{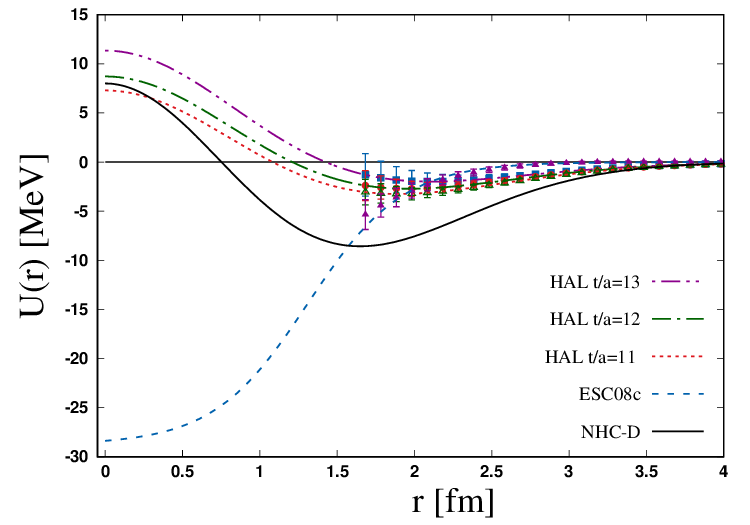}
		\caption{ The spin- and isospin averaged single-folding potentials $U_{\Xi t}$ in Eq.~\eqref{eq:SF-method} as functions of distance between $ \Xi $ and triton. 
			The first three based on  HAL QCD   $ \Xi {\text -}N $ potentials at $ t/a=13 $ (dash-dot-dotted magenta line), $ t/a=12 $ (dash-dotted green line), and $ t/a=11 $ (dotted red line). 
			The last two  $U_{\Xi t}$, i.e., ESC08c (dashed blue line) and NHC-D (solid black line) are constructed by using the Nijmegen extended soft-core and hard-core models, respectively.  The folding NHC-D potential is taken from Ref.~\cite{myint1994} as given by Eq.~\eqref{eq:U_Xi-t}. The fit range is taken to be $ r\gtrsim1.6 $ fm.
			\label{fig:XiT_pot}			
			}
	\end{figure*}
	
	According to the behaviour of the obtained single-folding potentials $U_{\Xi t}\left(r\right)$ in Fig.~\ref{fig:XiT_pot},
	 two different analytic functional forms is chosen for fitting.  
	The $ U_{\Xi t} $ potentials obtained based on the HAL potentials are fitted by the three-Gaussian functions as given by Eq.~\eqref{eq:U_Xi-t}, 
    and the other one built based on the ESC08c model is fitted by the following Wood-Saxon form 
    (inspired by common Dover-Gal model of potential~\cite{dover1983}) 
		
	\begin{equation}
		V_{\Xi t}^{fit}\left(r\right)=-V_{0}\left[1+\exp\left(\frac{r-R_{c}}{t}\right)\right]^{-1} , \label{eq:ws-fit}
	\end{equation}		
	where $ V_{0} $ is depth parameter, $ R_{c} = r_{c} A^{1/3} $ the radius of the nucleus		
	(with the mass number $ A = 3 $ for triton) and $ t $ the surface diffuseness.   
	In a nucleus, the $ R_{c} $ is measured from the center to a point where the density falls to around half of its value at the center.  
	The WS fit for the $\Xi t $ potential is constructed for the interval $r\gtrsim1.6$ fm~\cite{filikhin2024folding}. 
	This range is chosen according to the rms radii of triton from experimental measurements.

	The binding energy and  scattering observables for  $ \Lambda\left(\Xi\right){\text -}t $ systems are obtained by solving two-body Schr\"{o}dinger equation using the fitted $U_{Yt}\left(r\right)$ potentials (plotted in Figs.~\ref{fig:LambT_pot} and ~\ref{fig:XiT_pot}) as the input.
	In our numerical calculations, the mass of $ \Lambda $,  $ m_{\Lambda}=1115.683 $ MeV  is taken from the PDG~\cite{10.1093/ptep/ptaa104}, the mass of triton $ m_{t}=2808.921 $ MeV is from CODATA~\cite{mengCPC} and the mass of $\Xi$ is considered to be $m_{\Xi}=1318.07$ MeV. 
	The results for scattering length, effective range, and binding energy of the $ \Xi t $ are given in Table~\ref{tab:para-XiT}. 	
	We found that none of the potentials of the HAL $ t/a = 11, 12 $ and $ 13 $ support bound states. 
	Only for the NHC-D and ESC08c models, we have a possibility of a shallow bound state with the binding energies of $ 0.14 $ and $ 0.30 $ MeV with respect to the $ \Xi + t $ threshold.

\begin{figure*}[hbt!]
		\centering
				\centering
		\includegraphics[scale=0.64]{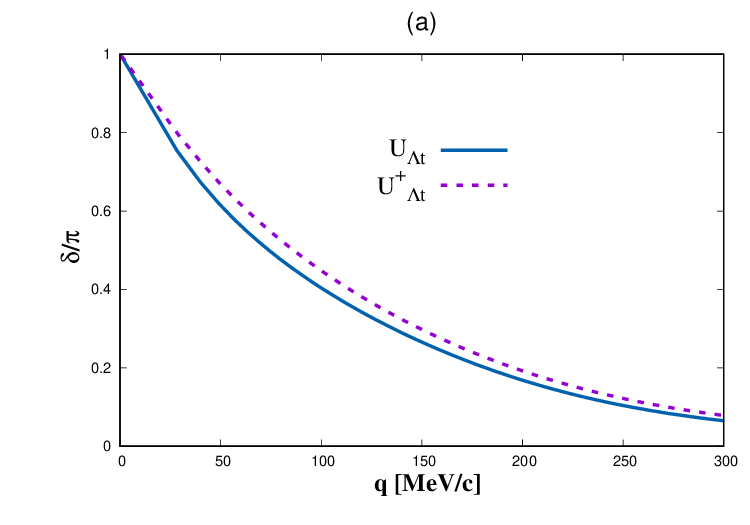} \includegraphics[scale=0.64]{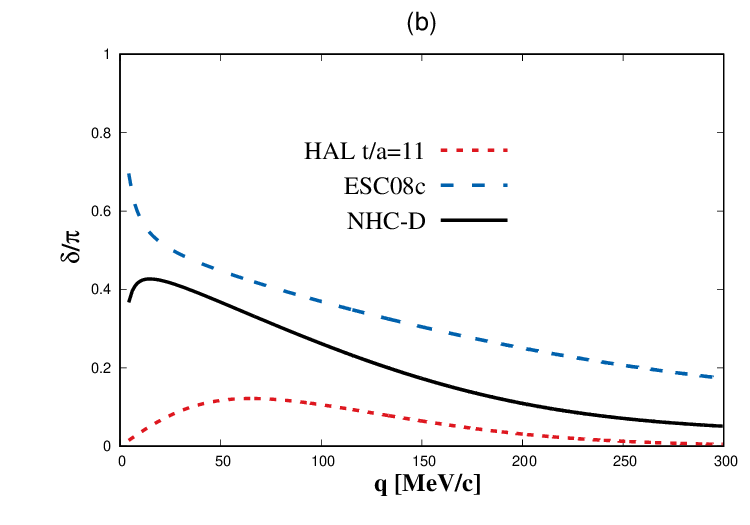}
		\caption{ 
			The normalized phase shifts as functions of the relative momentum $ q $ (a) for $ \Lambda t $  using $ U_{\Lambda t} $ (solid blue line), $ U_{\Lambda t}^{+} $ (dotted violet line); and (b) for $ \Xi t $ employing $ U_{\Xi t} $ based on the  HAL QCD $ t/a=11 $ (dotted red line), the ESC08c (dashed blue line) and NHC-D (solid black line) model of potentials. 
			\label{fig:phase_lt}}
	\end{figure*}

	Moreover, the relevance  $ Yt$ phase shifts $  \left(\delta/\pi\right) $  as functions of  
	the relative momentum ($ q=\sqrt{2\mu E} $ where $ \mu $ is the reduced mass of $ Yt $ system) is shown in Fig.~\ref{fig:phase_lt}. 
	The obtained phase shifts shows attractive behavior for all interactions.  
	Low-energy part of $ Yt $ phase shifts in Fig.~\ref{fig:phase_lt} defines scattering length $ a_{0} $ and effective range $r_{0}$  employing the effective range expansion (ERE) formula,		
	\begin{equation}
		q \cot\delta=-\frac{1}{a_{0}}+\frac{1}{2}r_{0}q^{2}+\mathcal{O}\left(q^{4}\right).\label{eq:ERE}
	\end{equation}		
	The numerical results for $ U_{\Lambda t}$ in  Eq.~\eqref{eq:U_lambdat},  
	$ U_{\Lambda t}^{+} $ in  Eq.~\eqref{eq:Uplus_lambdat} and $ \Xi t $ folding potential in Eq.~\eqref{eq:U_Xi-t}, are listed in table~\ref{tab:ltriton}.

	Hiyama et al. in Ref.~\cite{Hiyama2020} calculated the binding energy of $ NNN \Xi $ states  by a high precision variational approach, the Gaussian Expansion Method. They found for $ \Xi NNN $ system in ESC08c potential, the state in $ \left(T,J^{\pi}\right)=\left(0,0^{+}\right) $ is unbound with respect to $ \Xi+{}^{3}H/^{3}He $ threshold, while the states in $ \left(T,J^{\pi}\right)=\left(0,1^{+}\right),\left(1,0^{+}\right) $ and $ \left(1,1^{+}\right) $ are bound by $ 10.20,3.55 $ and $ 10.11 $ MeV, respectively. While with HAL QCD $ \Xi N $ potential they found that a possibility of a shallow bound state with the binding energies of $ 0.63\left(t/a=11\right),0.36\left(t/a=12\right),0.18\left(t/a=13\right) $ MeV with respect to the $ \Xi+{}^{3}H/^{3}He $ threshold.
	Since we used the spin- and isospin averaged $ \Xi N $ interaction to build the $ \Xi t $ potential, our results are some how different with GEM methods~\cite{Hiyama2020}, but the general behaviours are in agreement.
	\begin{table}[hbt!]
		\caption{
			The fit parameters of $\Xi t$ potentials, $ c_{i} $ (in MeV) and $ d_{i} $ (in fm) in Eq.~\eqref{eq:U_Xi-t}; 
			 $ V_{0} $ (in MeV), $ r_{c} $ (fm) and $ t $ (in fm) in Eq.~\eqref{eq:ws-fit}, 
			and the corresponding low-energy parameter, scattering length $ a_{0} $ (in fm), effective range $r_{0}$ (in fm) and binding energy $B_{\Xi t}$ (in MeV) are given for NHC-D, ESC08c and HAL QCD potentials.
			 The binding energy by including Coulomb interaction is given between parentheses, it implies that $ \Xi t $ system could be a Coulomb-assisted bound state~\cite{Hiyama2020,hiyamaPRC22}. 
			\label{tab:para-XiT}}. 	
		\begin{tabular}{c|ccccccccc}
			\hline
			\hline 
			Model-3G & $c_{1}$&$ d_{1}$&$c_{2}$&$d_{2}$&$c_{3}$&$d_{3}$& $a_{0}$&$r_{0}$& $B_{\Xi t}$ \\			
			\hline 		
			NHC-D   & $ 103.67$& $ 1.362 $ & $ -51.57 $ & $ 1.568 $ & $ -44.10 $ &$ 1.610 $& $-114.0 $& $3.1  $&-$\left(0.14\right)$\\
			HAL $ t/a=11 $&$29.3$&$0.56$&$-20.5$&$0.32$&$-1.5$&$0.17$&$-2.2$&$6.1$&-\\
			HAL $ t/a=12 $&$29.2$&$0.56$&$-19.1$&$0.33$&$-1.4$&$0.17$&$-1.6$&$7.5$&-\\
			HAL $ t/a=13 $&$29.2$&$0.56$&$-16.7$&$0.33$&$-1.2$&$0.17$&$-1.0$&$11.5$&-\\
			\hline \hline
			WS & \multicolumn{2}{c}{$ V_{0} $}  & \multicolumn{2}{c}{$ r_{c} $}& \multicolumn{2}{c}{$ t $}& $a_{0}$&$r_{0}$&  $B_{\Xi t}$ \\
			 \hline
		ESC08c	& \multicolumn{2}{c}{$ 28.8 $}  & \multicolumn{2}{c}{$ 0.90 $}& \multicolumn{2}{c}{$ 0.3 $}& $64.2$&$1.8$&- $\left(0.30\right)$ \\		   
			\hline 	\hline
		\end{tabular}
	\end{table}
	\section{Correlation function results and discussion}\label{sec:Two-particle-CF}
	The two-particle correlation produced in a collision~\cite{KOONIN197743,Pratt1986, PhysRevC.91.024916, OHNISHI2016294, cho2017exotic}  presents valuable information about the space-time evolution of the particle-emitting source and final state interactions entail hyperons~\cite{Fabbietti:2020bfg}.
	The $Yt$ correlation function is given in terms of the two-particle
	distribution $N_{Yt}\left(\boldsymbol{k}_{Y},\boldsymbol{k}_{t}\right)$
	normalized by the product of the single particle distributions, $N_{Y}\left(\boldsymbol{k}_{Y}\right)N_{t}\left(\boldsymbol{k}_{t}\right)$,
	\begin{equation}
		C\left(\boldsymbol{k}_{Y},\boldsymbol{k}_{t}\right)\equiv\frac{N_{Yt}\left(\boldsymbol{k}_{Y},\boldsymbol{k}_{t}\right)}{N_{Y}\left(\boldsymbol{k}_{Y}\right)N_{t}\left(\boldsymbol{k}_{t}\right)}, \label{eq:Cq1q2}
	\end{equation}
	where the $\boldsymbol{k}_{Y}$ and $\boldsymbol{k}_{t}$ are
	the momentum of particle Y and triton. 
	In the experiments, correlation function is measured by 
	\begin{equation}
		C\left(q\right)=\mathcal{N}\frac{A\left(\textbf{q}\right)}{B\left(\textbf{q}\right)}=\int4\pi r^{2}drS\left(\boldsymbol{r}\right)\left|\Psi_{Yt}^{\left(-\right)}\left(\boldsymbol{r},\boldsymbol{q}\right)\right|^{2}, \label{eq:kp}
	\end{equation}
	that is equivalent to Eq.~\eqref{eq:Cq1q2}. Where 
	$\textbf{q}=\left(m_{Y}\textbf{k}_{Y}-m_{t}\textbf{k}_{t}\right)/\left(m_{Y}+m_{t}\right)$
	is the relative momentum, and $A\left(\textbf{q}\right)$ is the distribution
	of $q$ with both particles from the same event, $B\left(\textbf{q}\right)$
	is for two particles from different events, and $\mathcal{N}$ is
	the normalization factor. 
	The right-hand side of Eq.~\eqref{eq:kp} is known as
	by the Koonin-Pratt (KP) formula~\cite{OHNISHI2016294}.
	$ S\left(r\right)=\exp\left(-r^{2}/4R^{2}\right)/\left(2\sqrt{\pi}R\right)^{3} $ with source size $ R $ 
	is known as the source function, which defines the distribution of the relative distance of particle pairs, 
	and is assumed to be spherical and static Gaussian in my calculations.
	$\Psi_{Yt}^{\left(-\right)} $ is the relative wave function of the particle pairs that can be obtained straightforwardly by solving a two-body Schr\"{o}dinger equation for a given  $ Yt $ potential.	
	
\begin{figure*}[hbt!]
		\centering
		\includegraphics[scale=0.64]{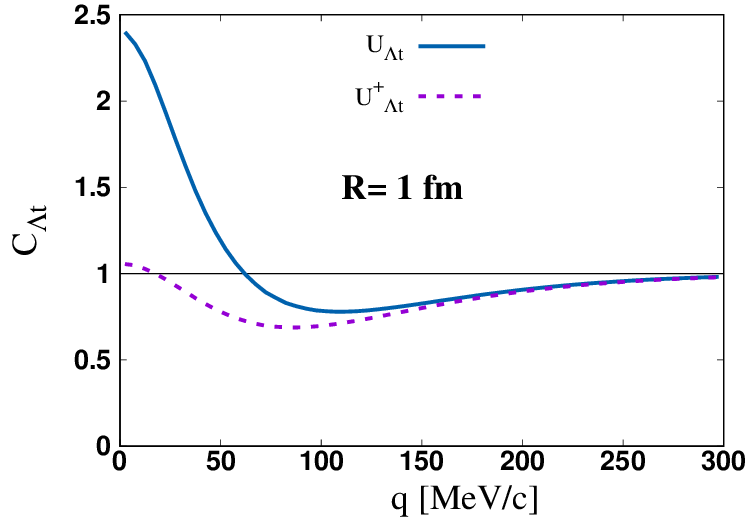} \includegraphics[scale=0.64]{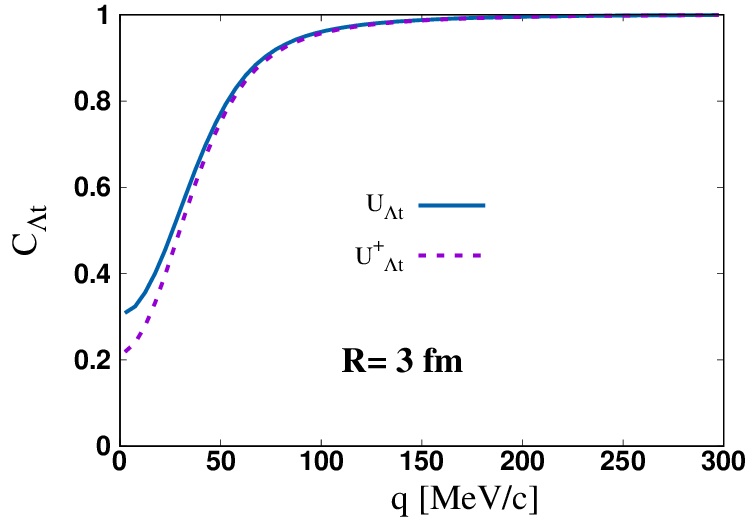} 
		\includegraphics[scale=0.64]{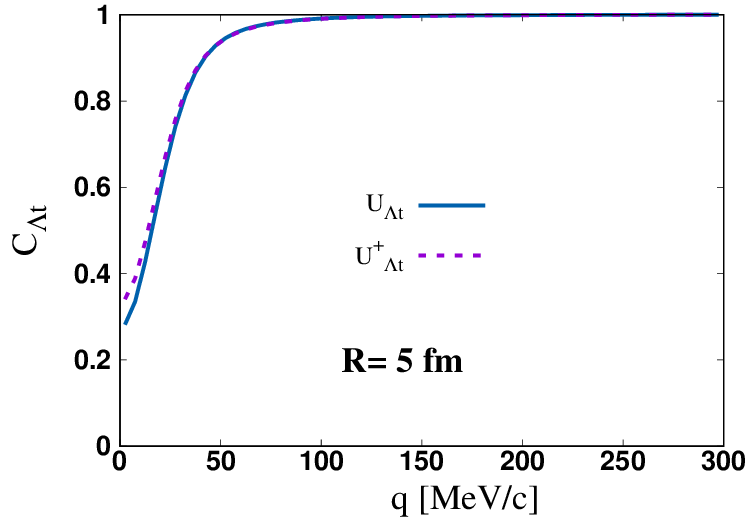} 
		\caption{
			The spin averaged $  \Lambda t $ correlation functions using $ U_{\Lambda t} $ (solid blue line), $ U_{\Lambda t}^{+} $ (dotted violet line) for three different source sizes: $ R=1,3 $ and $ R=5 $ fm.
			\label{fig:lambdaTriton_cq_kp_R}}	
	\end{figure*}

	The $ \Lambda t $ correlation functions for three different source sizes, 	$ R = 1, 3 $ fm and $ 5 $ fm  
	using $ U_{\Lambda t} $ and $ U_{\Lambda t}^{+} $ are shown in Fig.~\ref{fig:lambdaTriton_cq_kp_R}.
	The sizes of the sources have been chosen according to the sizes that are usually people used in
	the exploration of two-hadron correlation functions	~\cite{jinno2024femtoscopic, kamiya2024, etminan2024omegaAlpha}.	
	The particular dip shape can be observed at small source size $ R = 1 $ fm, that is conventional in the bound state of system near the threshold~\cite{jinno2024femtoscopic}. 	
	Moreover, Fig.~\ref{fig:lambdaTriton_cq_kp_R} for $ R = 1 $ fm shows the results for all cases of two potentials are different at low momentum $ q\lesssim100 $ MeV/c.		
	According to Fig.~\ref{fig:LambT_pot} for $ R = 1 $, in the low momentum
	region  fm,  the $ U_{\Lambda t}$ potential model, is more attractive than $ U_{\Lambda t}^{+} $ potential, thus it gives enhancement of $ C_{\Lambda t} $, i.e., The $ C_{\Lambda t} $ function suppressed due to the strengthening of the repulsive core part in $ U_{\Lambda t}^{+} $ potential. 
	Nevertheless, with the increase of the source size, the difference between 
	the $ C_{\Lambda t}\left(q\right) $s decreases until they are almost identical for $ R = 5 $ fm.
	As a consequence, the future measurements of $ \Lambda t $ correlation functions in the relatively low momentum region with small radius of the sources, can be used as a probe to study the $ \Lambda t $ interaction in dense matter.
	
\begin{figure*}[hbt!]
		\centering
		\includegraphics[scale=0.64]{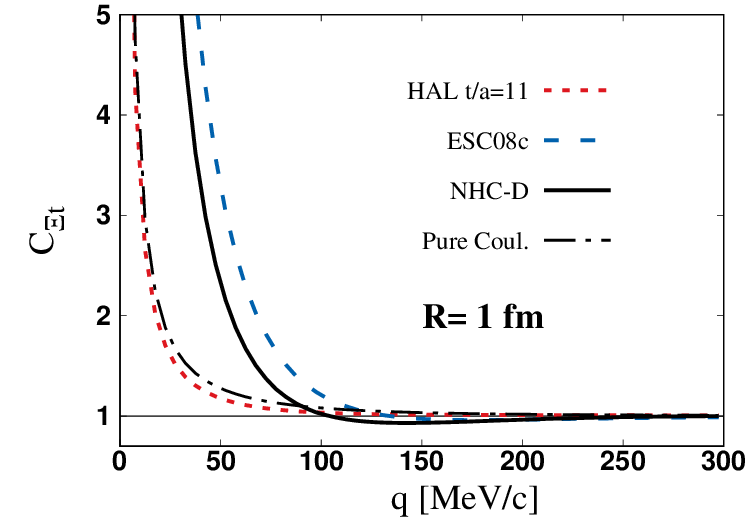} \includegraphics[scale=0.64]{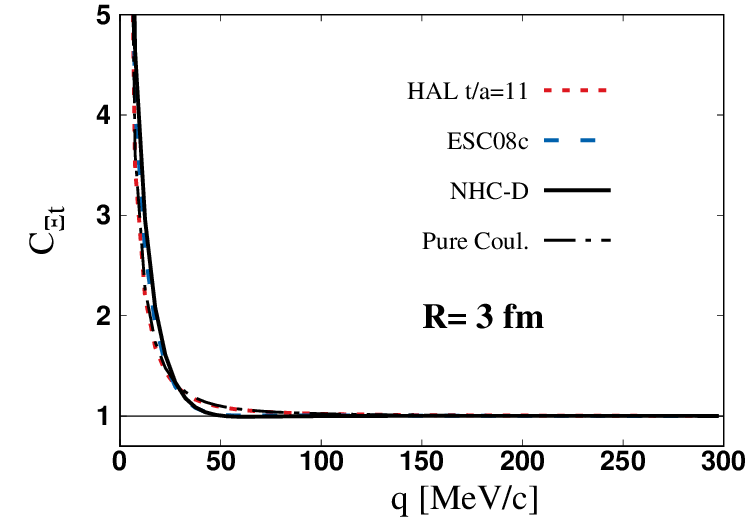} 
		\includegraphics[scale=0.64]{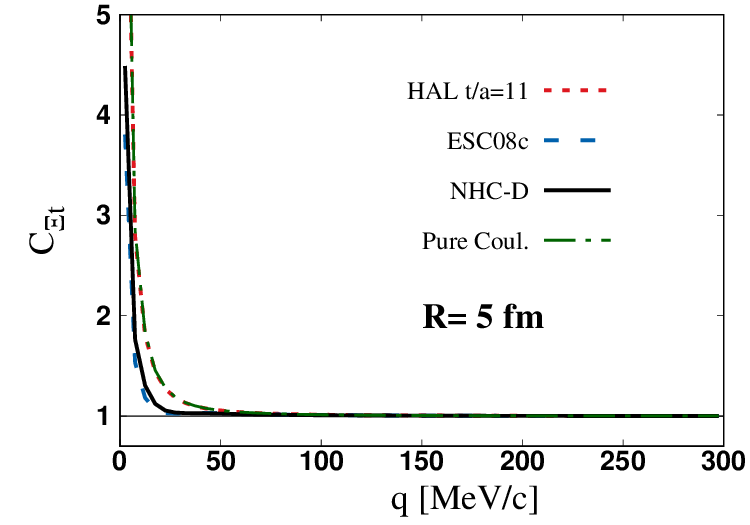} 
		\caption{
			The spin- and isospin averaged $ \Xi^{-} t $ correlation functions using 			
				the HAL QCD $ t/a=11 $ (dotted red line), ESC08c (dashed blue line) and NHC-D (solid black line) $ U_{\Xi t} $ potentials for three different source sizes: $ R=1, 3 $ and $ R=5 $ fm, when the Coulomb interaction is switched on. 
			For comparison, the pure Coulomb result, where the strong interaction is switched off, is presented by the dash-dotted green line.		 
			\label{fig:XiTriton_cq_kp_R}}	
	\end{figure*}

	The $ \Xi t $ correlation functions using 			
	the HAL QCD $ t/a=11 $, ESC08c and NHC-D models of potentials for the three different source sizes are presented in  Fig.~\ref{fig:XiTriton_cq_kp_R}. 
	In this case, the Coulomb potential is also included.
	It is seen a  substantial enhancement in $ \Xi t $ correlation at low momentum region. 
	That implies a relatively large scattering length (effective range) value for ESC08c, NHC-D (HAL) models of potentials (see table~\ref{tab:para-XiT}) and  the Coulomb attraction.
	In order to see the effect of strong interaction, also, the pure Coulomb potential is plotted in Fig.~\ref{fig:XiTriton_cq_kp_R}. 	
	However, with good measurement resolution, it might be possible to recognize different potentials with a correlation function with source size, $ R = 1-3 $ fm.	
	Specifically,  $ \Xi^{-} t $ correlation functions are enhanced at the small source and suppressed at the significant source analogous to the pure Coulomb calculation. 
	Moreover, the difference between $ \Xi t $ correlation function and pure Coulomb result,
	which is embodied as a dependency of the correlation function on the size of the source, 
	it implicitly implies the existence of a Coulomb-assisted bound state~\cite{kamiya2024}.

	\section{Summary and conclusions\label{sec:Summary-and-conclusions}}	
	In this exploratory study, $ \Lambda $ and $ \Xi{\text -}$triton correlation functions in heavy-ion collisions were predicted. They can be measured in the STAR detector at RHIC~\cite{AbdallahPLB2022} by employing the femtoscopy technique. 
	
	For $\Lambda t$ two-body potential, Kurihara's isle-type potential~\cite{KuriharaPRC1985}, 
	that is extracted from Dalitz's hard core $\Lambda N$ interaction was used. 
	The strength of $ \Lambda t $ potential was tuned in  such a way as to return
	the experimental ground state energy of $_{\Lambda}^{4}H$~\cite{JURIC19731, YamamotoPRL2015}.
	In addition, since the new measurements show a relatively significant increase  in the $\Lambda$ binding energy of the hypertriton and $_{\Lambda}^{4}H$ hypernuclei~\cite{EsserPRL2015, SchulzNPA2016, AbdallahPLB2022},
	I also performed the calculations with an increased strength of $ \Lambda t $ potential by about twenty percent.

		Moreover in the case of $\Xi t$ two-body potential, because there is no experimental measurement on the $ \Xi{\text -}t $ interaction yet, I presented 
	a three Gaussian and a  Woods-Saxon type form for $\Xi t$ potential based on the first principles HAL QCD and  ESC08c model of $\Xi N$ interactions, respectively.
	The $\Xi t$ potentials were obtained in the SFP approach by using the corresponding spin- and isospin averaged $\Xi N$ interactions. 
	In addition, another $\Xi t$ potential built based on  Nijmegen hard-core model D (NHC-D) $\Xi N$ potential which is taken from the literature~\cite{myint1994}, was examined. 	
	The ESC08c and NHC-D models of $\Xi N$ interactions returned the large scattering length without a bound state (a shallow Coulomb-assisted bound state) for the 
	$ \Xi t $ system.

	Employing these obtained $ Yt $ potentials, correlation functions were calculated  using the KP formula for three different source sizes, $ R=1,3 $ fm and $ 5 $ fm, 
	where the choice was motivated by values suggested by analyses of measurements of the two-hadron correlation function in $ pp $ collisions and heavy ion collisions~\cite{jinno2024femtoscopic,kamiya2024}.
	The numerical results  suggested that, with good measurement resolution, it might be possible to recognize different potentials with a correlation function at relatively small source sizes, i.e., $ R = 1-3 $ fm. 
	It is remarkable that the KP formula in Eq.~\eqref{eq:kp} is accurate under the condition that two correlated particles can be treated as point-like objects. 
	The  source size of the composite particle $ t $ should  be bigger than those with single hadron emissions,  
	for the reason that, there is a probability of $ t $ particle formation by the coalescence of three nucleons emitted
	from the fireball~\cite{mrowczynski2019hadron, bazak2020production, StanislawPRC2021}.
	Therefore, we rather deal with a four-body scattering problem of a three nucleons and a hyperon.
	In the other hand, a formation of triton  and production of $ \Lambda t $ correlation arise at same time. 	
	Essentially, this mechanism must be taken into account in future studies and it is beyond the scope of this paper.	
	
	In conclusion, the numerical results of this work suggest that measurements of the $ \Lambda t $ momentum correlation function
	in heavy-ion collisions~\cite{AbdallahPLB2022, kozhevnikova2024lighthypernuclei} look indeed very promising.
	Particularly, collected data by the STAR detector from Au+Au collisions at a center-of-mass energy of $ \sqrt{s_{NN}}=3 $ GeV  
	provide an opportunity to study the $ \Lambda t $ correlation function.
		
	\section*{Acknowledgement}
	I am grateful to the authors and maintainers of "\textit{Correlation Analysis tool using the Schr\"{o}dinger Equation}" (CATS)~\cite{cats}, a modified version of which is used for calculations in this exploratory study.	
	Discussions during the long-term workshop, HHIQCD2024 at Yukawa Institute for Theoretical Physics (YITP-T-24-02), 
	aroused in me the motivation for this work.

	
	\bibliography{Refs.bib}
	
\end{document}